# Title page

**Title**: Symmetry-compatible angular momentum conservation relation in plasmonic vortex lenses with rotational symmetries

Jie Yang[1†], Pengyi Feng[2†], Fei Han[3,4], Xuezhi Zheng[1*], Jiafu Wang[5*], Zhongwei Jin[6], Niels Verellen[4,3], Ewald Janssens[3], Jincheng Ni[7], Weijin Chen[7], Yuanjie Yang[8], Anxue Zhang[5], Benfeng Bai[2*], Chengwei Qiu[7*], and Guy A E Vandenbosch[1]

Jie Yang
[1]WaveCoRE research group, KU Leuven, Kasteelpark Arenberg 10, BUS 2444, Leuven B-3001, Belgium
Email: jyang@esat.kuleuven.be

Pengyi Feng
[2]State Key Laboratory of Precision Measurement Technology and Instruments, Department of Precision Instrument, Tsinghua University, Beijing 100084, China
E-mail: fengpy21@mails.tsinghua.edu.cn

Fei Han
[3]Quantum Solid-State Physics, Department of Physics and Astronomy, KU Leuven, 3001 Leuven, Belgium
[4]IMEC, 3001 Leuven, Belgium

Xuezhi Zheng*, corresponding author
[1]WaveCoRE research group, KU Leuven, Kasteelpark Arenberg 10, BUS 2444, Leuven B-3001, Belgium
Email: xuezhi.zheng@esat.kuleuven.be

Jiafu Wang*, co-corresponding author
[5]Xi'an Jiaotong University, Xianning West Road 28, Xi'an City, Shaanxi province 710049, China.
Email: wangjiafu1981@126.com

Zhongwei Jin
[6]College of Optical and Electronic Technology, China Jiliang University, Hangzhou, 310018, China

Niels Verellen
[4]IMEC, 3001 Leuven, Belgium
[3]Quantum Solid-State Physics, Department of Physics and Astronomy, KU Leuven, 3001 Leuven, Belgium

Ewald Janssens
[4]Quantum Solid-State Physics, Department of Physics and Astronomy, KU Leuven, 3001 Leuven, Belgium

Jincheng Ni
[7]Department of Electrical and Computer Engineering, National University of Singapore, 4 Engineering Drive 3, Singapore 117583, Singapore

Weijin Chen
[7]Department of Electrical and Computer Engineering, National University of Singapore, 4 Engineering Drive 3, Singapore 117583, Singapore

Yuanjie Yang
[8]School of Physics, University of Electronic Science and Technology of China, Chengdu 611731, China

Anxue Zhang
[5]School of Information and Communications Engineering, Xi'an Jiaotong University, Xianning West Road 28, Xi'an City, Shaanxi province 710049, China.

Benfeng Bai
[2]State Key Laboratory of Precision Measurement Technology and Instruments, Department of Precision Instrument, Tsinghua University, Beijing 100084, China
E-mail: baibenfeng@tsinghua.edu.cn

Chengwei Qiu
[7]Department of Electrical and Computer Engineering, National University of Singapore, 4 Engineering Drive 3, Singapore 117583, Singapore
E-mail: chengwei.qiu@nus.edu.sg

Guy A E Vandenbosch
[1]WaveCoRE research group, KU Leuven, Kasteelpark Arenberg 10, BUS 2444, Leuven B-3001, Belgium.
Email: guy.vandenbosch@kuleuven.be

† These authors contributed equally: Jie Yang and Pengyi Feng.

**Abstract:** Plasmonic vortex lenses (PVLs), producing vortex modes, known as plasmonic vortices (PVs), in the process of plasmonic spin-orbit coupling, provide a promising platform for the realization of many optical vortex-based applications. Very recently, it has been reported that a single PVL can generate multiple PVs. This work exploits the representation theory of finite groups, reveals the symmetry origin of the generated PVs, and derives a new conservation relation based on symmetry principles. Specifically, the symmetry principles divide the near field of the PVL into regions, designate integers, which are the topological charges, to the regions, and, particularly, give an upper bound to the topological charge of the PV at the center of the PVL. Further application of the symmetry principles to the spin-orbit coupling process leads to a new conservation relation. Based on this relation, a two-step procedure is suggested to link the angular momentum of the incident field with the one of the generated PVs through the symmetries of the PVL. This theory is well demonstrated by numerical calculations. This work provides an alternative but essential symmetry perspective on the dynamics of spin-orbit coupling in PVLs, forms a strong complement for the physical investigations performed before, and therefore lays down a solid foundation for flexibly manipulating the PVs for emerging vortex-based nanophotonic applications.

**Introduction**

Since the pioneering work of Allen et al.[1], it is known that electromagnetic (EM) waves can carry not only spin angular momentum (SAM) in the form of circular polarizations, but also orbital angular momentum (OAM) in the form of vortex modes. The vortex mode carrying OAM has a spiral-type phase singularity where the intensity is zero and the phase is undetermined. The order of the phase singularity, also known as the topological charge of the vortex mode, describes how many full phase variations (from 0 to $2\pi$) the wavefront experiences during the propagation within a wavelength. In this work, we will use the following three nomenclatures: the order of the phase singularity, the topological charge, and the number of full phase variations, interchangeably. The OAM can be evaluated as $l\hbar$ where $l$ is the topological charge and $\hbar$ is the reduced Planck constant. Different from the well-known two-state SAM, OAM brings a new and theoretically unbounded degree of freedom for engineering EM waves and therefore inspires numerous novel physical applications in super-resolution microscopy[2], classical[3-5] and quantum communications[6-7], optical manipulation[8-9], and quantum information processing[10] etc.

At optical frequencies, an intriguing approach to generating OAM is by using plasmonic nanostructures like plasmonic vortex lenses (PVLs). In this context, the vortex modes are aliased as "plasmonic vortices" (PVs). The PV is supported by surface plasmon polaritons (SPPs) which are evanescently confined around the structures[11-12]. Owing to the elegant physical properties such as spin-dependent excitation and subwavelength field localization, the PVL provides an ideal platform for studying spin-orbit coupling[11-18] and vortex light-matter interaction[19-20], and realizing many exciting applications like highly integrated on-chip OAM sources[15] and optical tweezers[9,21]. It has been well established that the PVs can be generated by the PVLs in the process of spin-orbit coupling[11-19]. The adopted PVL usually holds $M$-fold discrete rotational symmetries (DRSs) and is under the illumination of a circularly polarized vortex beam carrying OAM $l\hbar$ and SAM $\sigma\hbar$ ($\sigma = \pm 1$, where the positive/minus sign corresponds to the left/right circular polarization). The generated PV is firstly demonstrated to carry the topological charge of $(l + \sigma) + M$[13,15]. This formula provides a conservation relation that links the angular momentum of the incident beam with the one of the generated PV through the order of rotational symmetries of the PVL. Although this conservation relation can well explain the order of the generated plasmonic vortices, it cannot fully explain the behavior of the generated PVs in the near field region of a PVL where complex interference patterns appear other than the phase singularities with the order of $(l + \sigma) + M$[15]. For this, the latest work of Yang et al. has demonstrated that there exist multiple deuterogenic PVs in a single PVL. There, besides the conventional topological charge of $l + \sigma + M$ [16], multiple PVs are found in the near field region of a PVL and the PVs are demonstrated to carry the topological charges of $l + \sigma + Mq$ where $q$ is an integer[14,16]. It is also believed that the PV around the central region of the PVL carries a topological charge $l + \sigma$ (see Discussion and Conclusions in [16]). Although this generalized conservation relation well explains the order of the generated multiple plasmonic vortices, there is still a potential incompatibility with the symmetry principles. That is, since the OAM and the SAM of the incident light can be arbitrary, the topological charge $l + \sigma$ of the PV in the central region looks unbounded. However, in many electronic systems holding DRSs such as chromophore arrays[22-24] and nanopillar arrays[25-26], the topological charges of the vortex modes do not exceed an upper bound which equals the nearest-lower integer of (*M*-1)/2, i.e., $\left[\frac{M-1}{2}\right]$. Such an upper bound is defined by the representation (rep.) theory of finite groups.

This observation motivates us to study how symmetry principles work in the process of spin-orbit coupling in PVLs, by which it can be promising to reveal the symmetry origin of the PVs in the near field region of a PVL. To this end, we harness the rep. theory of finite groups to study the effects of symmetries on the interaction of light with PVLs. Firstly, we study the topological properties of the PVs in a PVL without considering a specific incident light. We reveal that the rep. theory divides the near field of the PVL into circular and ring-like regions, and assign each region an integer, which is topological charge of the region. Especially, for the inner-most region, i.e., the central region, there exists an upper bound for the topological charges, i.e., $\left[\frac{M-1}{2}\right]$, where the squared bracket rounds the number in the bracket to the nearest-lower integer. Secondly, we consider the interaction of the PVL with an incident light, i.e., the process of spin-orbit coupling in PVLs. There, we propose an angular momentum conservation relation for the PVLs with DRSs. This relation is compatible with the symmetry principles, i.e., the rep. theory of finite groups. Based on this, we also suggest a two-step procedure that links the angular momentum of the incident light with the one of the generated PVs, through the order of the symmetries of the PVL. Finally, we confirm the theory by numerical simulations. In this work, the time convention $e^{-i\omega t}$ is used and will be suppressed in following discussion.

**Results**

*Key Observations for the $C_8$ Group and their symmetry origins* Without loss of generality, a PVL with 8-fold rotational symmetries is first considered. In this case, the rotational symmetries form a cyclic group, $C_8$ (see Figure 1(a)). The PVL, which consists of eight spiral strips, is a complementary structure of the conventional PVLs[15-16]. The strips are described as $\rho=\rho_1+8\varphi\lambda_1/2\pi-(a-1)\lambda_1$, where $\rho_1=1.5\lambda_1$ and $a=1,\ldots,8$. The width of the strip is $0.25\lambda_1$ (See Figure 1(a) for the definitions of $\rho_1$ and $\lambda_1$). Based on the rep. theory of finite groups[27-28], the $C_8$ group defines 8 irreducible representations (irreps) and, accordingly, 8 possible categories of electric field distributions that can be radiated (or scattered) by the PVL (see SI S1 for the theory). The electric field distributions are numerically evaluated and shown in Figure 1(b)-(i) (see computational details in the SI S2). Each irrep and its corresponding electric field distribution are indexed by an integer, $j$. As an abelian group, the $C_8$ group has eight irreps[27-28]. As a result, $j$ ranges from 0 to 7. For each index $j$, we first focus on the circular region enclosed by the white dashed circle. Here, there are **two key observations**. On the one hand, except for the case $j=4$, the electric field distributions of all other cases exhibit the features of the vortex modes. Specifically, in Figure 1(b)-(d) and (f)-(i), it is seen that the vortex modes carry the topological charges of -1, -2, -3, +3, +2, +1 and 0, respectively. In Figure 1(e), an octopole mode rather than a vortex mode is formed. On the other hand, the allowed topological charges, i.e., $l_s$'s, are bounded. Specifically, in Figure 1(b)-(d) and (f)-(i), for the $C_8$ group, an upper bound for $|l_s|$ of 3 is observed.

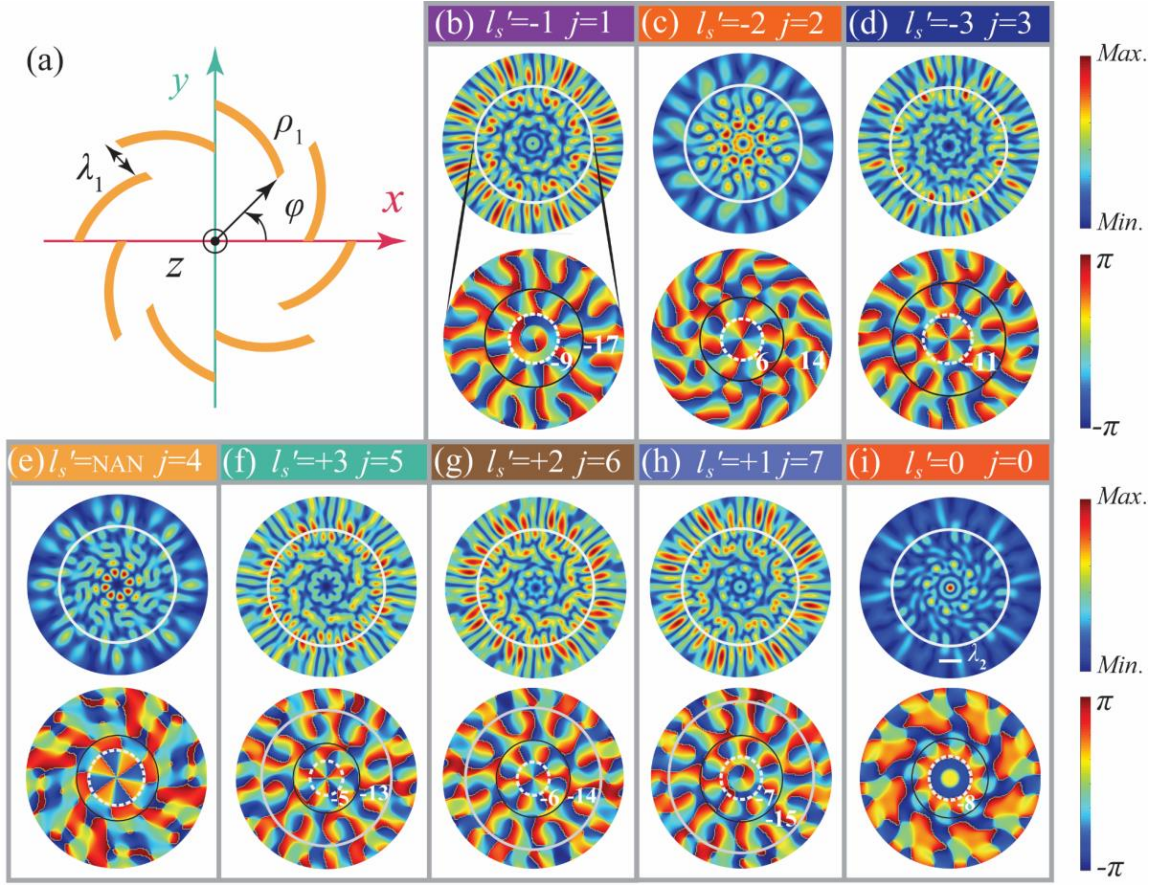

**Figure 1.** Illustration of the PVL with the $C_8$ group and the $z$ component of the electric fields on a cut plane above the structure ($z = 1.8\lambda_1$). In (a), the PVL with 8-fold rotational symmetries is presented. The strips of the PVL are assumed to be made of metals, e.g., Au. In (b)-(i), the magnitude and phase patterns of $E_z$ corresponding to the irreps are shown. Specifically, the upper panels in (b)-(i) plot the magnitude, while the lower panels in (b)-(i) do the phase. Each lower panel plots the phase distribution of the region encircled by the white solid line in the corresponding magnitude plot. In the lower panels, the inner-most circular regions, i.e., the central regions, are enclosed by the white dashed lines; the first ring-like regions are delineated by the white dashed lines and the black solid lines; the second ring-like regions, if exist, are delineated by the black solid lines and the grey solid lines. Each circular or ring-link region is marked by an integer (in white). The magnitude is coded by the color from blue to red to denote the smallest and the largest values; the phase is coded by the same color to denote the phase variations between $-\pi$ and $+\pi$.

Next, for each $j$, we focus on the ring-like regions outside the white dashed circles in Figure 1. ***Another key observation*** is that more full-phase variations exhibit in the outer region. In Figure 1(b) – (d) and (f) – (i), the number of full-phase variations is marked by a white number. This number is defined in such a way that: when the phase revolves around the phase singularity in a clockwise manner, the number is a negative integer, while, for the counterclockwise case, a positive integer. It is observed from the phase distributions in Figure 1(b)-(d) and (f)-(i) that, for the region between the white dashed line and the black solid line, the number of full phase variations is equal to the topological charge in the central region plus $\pm 8$ (see the white numbers in Figure 1); for the region outside the black solid line in Figure 1(b)-(d), and the region between the black solid line and the grey solid line in Figure 1(f)-(h), the number of full phase variations is equal to the topological charge in the central region plus $\pm 2 \times 8 = \pm 16$ (see the white numbers in Figure 1).

To exploit the symmetry origin of the above three key observations, we consider a simple dipole model. The use of the dipole model is motivated by the fact that, due to the volume equivalence principle[29-33], the scattered electric fields observed in Figure 1 can be seen as being generated by the induced electric currents flowing in the PVL; and the fact that the currents can be always seen as the superposition of point sources[29-33], i.e., electric dipoles. Consider eight electric dipoles located at points $\mathbf{r}'_0$-$\mathbf{r}'_7$ so that the dipole set (see

Figure 2) holds the same $C_8$ group (to be precise, it's the $D_8$ group, see SI S3 for more information) as the PVL in Figure 1 does. For the sake of simplicity, the harmonic dipoles oscillating at an angular frequency $\omega$ (related to $\lambda_1$ in Figure 1 (a)) are assumed to be oriented along the positive $z$ direction and are arbitrarily weighted. In details, at the point $\mathbf{r}'_m$, the $m^{th}$ dipole takes the moment $p_0(\mathbf{r}'_m)\hat{z}$ where $p_0(\mathbf{r}'_m) = \alpha_m p_0$ with $\alpha_m$ (a complex number) being the weight. According to the rep. theory of finite groups, any arbitrarily weighted dipole set can be decomposed as the superposition of eight orthogonal dipole sets. The eight orthogonal dipole sets correspond to the eight irreps of the $C_8$ group (see details in the SI S3). For each orthogonal dipole set, the magnitudes of the dipoles are the same (which will be omitted in the following discussions) and the phases of the dipoles are defined by the corresponding irrep of the $C_8$ group. In details, for the $j^{th}$ irrep, the weight of the $m^{th}$ dipole is $\varepsilon(j)^{-m}$, where $\varepsilon(j)$ is a function of $j$ and reads $e^{ij2\pi/M}$ with $M = 8$ (see Figure 2).

The electric field radiated by the dipole set corresponding to the $j^{th}$ irrep is $\mathbf{E}(\mathbf{r}) = \sum_{m=0}^{7} \overline{\mathbf{G}}(\mathbf{r},\mathbf{r}'_m) \cdot \varepsilon(j)^{-m} p_0 \hat{z}$, where $\mathbf{r}$ denotes an observation point and $\overline{\mathbf{G}}$ is the dyadic Green function[29-30,32]. Especially, the $z$ component of the radiated field is $E_z(\mathbf{r}) = p_0 \sum_{m=0}^{7} G_{zz}(|\mathbf{r} - \mathbf{r}'_m|)\varepsilon(j)^{-m}$. $G_{zz}$ is the $zz$ component of the dyadic Green function and is only dependent on the distance between an observation point $\mathbf{r}$ and a source point $\mathbf{r}'_m$.

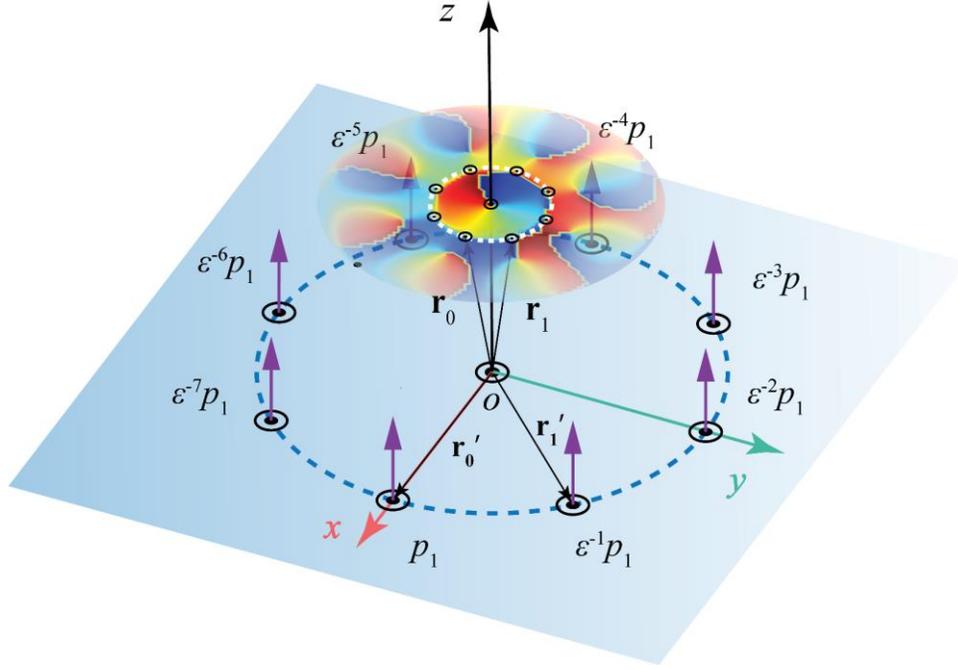

**Figure 2.** An illustration of the weighted dipole set and its radiated field. In the figure, the dipoles sit on the $xy$ plane (see the Cartesian coordinate system). Especially, $p_1 = \left(\frac{1}{8}\sum_{m=0}^{7}\alpha_m\right)p_0$; and $\varepsilon$ is short for $\varepsilon(j) = e^{ij2\pi/8}$ and, in this figure, $j$ is taken to be 1. Then, the phase distribution of a possible electric field configuration is illustrated over a circular region on a plane just above the $xy$ plane. The circular region is divided into two parts: the inner-most region encircled by a white dashed line and the region outside the line. Also, along the white dashed line, there are eight observation points marked by solid black circles. The observation points are so chosen that the geometric angle between two adjacent points is $\pi/4$. In the inner-most region, since the case $j = 1$ is considered, only one full phase variation is seen; while, in the outer region, 8 more full phase variations are observed. The blue and the red colors are used to cover the phase variation between $-\pi$ and $+\pi$.

Choose an observation plane parallel to the plane where the dipole set sits, as shown in Figure 2. On the one hand, consider the point at which the observation plane and the $z$ axis intersect. Since the distances between this field point and all the source points are equal, this point is a rotation center. At the point, the electric field is,

$$E_z = G_{zz} p_0 \sum_{m=0}^{7} \varepsilon(j)^{-m} = \begin{cases} 0, & j \neq 0 \\ 8 G_{zz} p_0, & j = 0 \end{cases}. \tag{1}$$

In Equation (1), the Green function $G_{zz}$ is extracted as common factor, since it is only a function of the distance. The equation suggests that when $j \neq 0$, the electric field at the rotation center is zero and is thus the phase singularity; when $j = 0$, the electric field is maximal, which corresponds to the trivial vortex mode with zero topological charge. On the other hand, consider eight field points on the observation plane (see Figure 2). The eight field points are chosen in such a way that two adjacent points are related by a 45-degree rotation. Since the Green function $G_{zz}$ is a function of the distance, only the relative position of a field point with respect to a source point is important. The $z$ component of the electric field at the field point $\mathbf{r}_0$ generated by the dipole at $\mathbf{r}_0'$ is the same as the one at the field point $\mathbf{r}_1$ generated by the dipole at $\mathbf{r}_1'$ in magnitude but is different by a $-j2\pi/8$ in phase. The same line of reasoning applies, when all eight dipoles are considered. As a result, there is always a $-j2\pi/8$ phase difference between the fields at two adjacent field points (see Figure 2 where $j = 1$ is taken as an example). This phase difference has *three* consequences. Firstly, it defines the total number of full phase variations along a closed path around the rotation center, i.e., the order of the phase singularity. Hence, the topological charge $l_s$ carried by the vortex mode is thus determined by $j$: (i) $j < 4$, $l_s = -j$; (ii) $j > 4$, $l_s = M - j$; (iii) $j = 4$, here the phase is $-\pi$ that leads to the weights jumping between +1 and -1, so that the field distribution corresponding to the 4th irrep is not a vortex mode. Secondly, since the irrep index $j$ can only take the integers between 0 and 7, this phase difference defines the **maximal** phase step between two adjacent field points, gives a **maximally** allowed number of full phase variations along the closed path, and hence puts **an upper limit** on the order of the phase singularity. For the $C_8$ group, the order is up to ±3. Thirdly, it is noted that $\varepsilon(j)$ is a periodic function of $8q$, i.e., $e^{i(j)2\pi/8} = e^{i(j+8q)2\pi/8}$, where $q$ is an integer. This relation implies that, in the most general case, the phase variation between two adjacent field points can be $-(j + 8q)\,2\pi/8$. That is, in general, $q$ full phase variations can be accommodated along the arc connecting two adjacent field points, which can be demonstrated by the regions between the white dashed lines and the black solid lines in the subplots of Figure 1. In this sense, the previous relation between the topological charge $l_s$ and the index of the $j^{th}$ irrep can be generalized as: (i) $j < 4$, $l_s = -j + 8q$ and (ii) $j > 4$, $l_s = M - j + 8q$. Physically, to accommodate $q$ full phase variations, the arc length must be multiples of the wavelength. As a result, it is natural that the central regions (encircled by the white dashed lines in Figure 1) implying small arc lengths can only exhibit bounded low-order topological charges, while the outer regions (outside the white dashed lines in Figure 1) implying large arc lengths do show higher order topological charges with differences of $8q$.

*Generalization to Arbitrary Finite Cyclic Groups* The above three consequences well echo *the three key observations* made at the beginning of the previous section. They can be immediately generalized to the case of PVLs with *M*-fold rotational symmetries

(which form the $C_M$ group). The topological charge $l_s$ of a vortex mode radiated by the PVL is related with the index of the $j^{th}$ irrep of the $C_M$ group,

$$l_s = \begin{cases} -j + Mq, & j < M/2 \\ M - j + Mq, & j > M/2 \end{cases}. \qquad (2)$$

In the above equation, $j$ ranges from 0 to $M - 1$ and, again, $q$ is an integer. When $M$ is even, there is an $(M/2)^{th}$ irrep. This irrep does not correspond to a vortex mode with the topological charge of $\pm(M/2)$ but a $M$-pole mode. Further, the above equation already includes the cases where $q$ full phase variations are present in the vortex modes. As a result, for a fixed irrep, e.g., the $j^{th}$ irrep, there can be infinitely many corresponding topological charges, since $q$ can be an arbitrary integer. We may introduce the concept of *modular* topological charges $l'_s$ to impose a one-to-one correspondence between the index of an irrep and the topological charge,

$$l'_s = l_s \bmod M = \begin{cases} -j, & j < M/2 \\ M - j, & j > M/2 \end{cases}. \qquad (3)$$

Since $j$ is an integer between 0 and $M - 1$, by Equation (3), the *modular* topological charge $l'_s$ is hence bounded. If $M$ is an odd number, the upper bound is $\frac{M-1}{2}$; if $M$ is an even number, the upper bound is $\frac{M}{2} - 1$. In short, the $C_M$ group permits *modular* topological charges $|l'_s| = 1, \cdots, \left[\frac{M-1}{2}\right]$, where the square bracket rounds $\frac{M-1}{2}$ to the nearest non-negative integer towards zero.

*Physically*, the unprimed topological charge $l_s$ divides the near field of a PVL into regions and designates each region an integer (see Figure 1(b)-(d) and (f)-(i)). In details, the central region is characterized by a PV carrying a topological charge $l'_s$ and the outer regions are by PVs carrying topological charges $l'_s + Mq$. Therefore, in this picture the mode is seen as a superposition of multiple PVs. In contrast, the *modular* topological charge $l'_s$ treats the vortex mode as a whole and associates the entire mode with a unique integer. Therefore, there is always a one-to-one correspondence amongst the *modular* topological charge $l'_s$, the index of an irrep $j$ and a vortex mode supported (scattered) by a PVL.

*A Conservation Relation and Verifications* The two previous sections focus on the *eigen-properties* of PVLs without considering a specific incident light. In this section, the role of the incident light carrying an orbital angular momentum of $l\hbar$ and a spin angular momentum of $\sigma\hbar$ is exploited. By the rep. theory of finite groups (see derivations in the SI S4), a conservation relation is found to link the orbital and the spin angular momenta of the incident light, i.e., $l$ and $\sigma$, with the index of an irrep, i.e., $j$, through the fold of rotational symmetries, i.e., $M$, held by a PVL,

$$(l+\sigma)+j = Mq. \tag{4}$$

It is essential to note that $j$ be an integer between $0$ and $M-1$. As discussed previously, the index $j$ corresponds to a vortex mode radiated (scattered) by the PVL. Hence, Equation (4) provides an essential link amongst the incident field, the scattered field and the symmetries of the scatterer. Since the derivations (see SI S4) are independent from the context of plasmonic vortices, this relation can be used for generic electromagnetic systems holding rotational symmetries. Further, one may be inclined to directly substitute Equation (2) to Equation (4). This substitution may lead to $l_s = (l+\sigma) + Mq$. This relation is the well-accepted angular momentum conservation relation[16]. It is also believed that $l+\sigma$ gives the topological charge at the center of a PVL (see Discussion and Conclusions in [16]). Since the incident light can carry OAM and SAM up to arbitrary orders, the topological charge at the center of a PVL is seemingly unbounded. However, extra attention needs to be paid to the substitution. Essentially, it is noted that $j$ be an integer bounded by 0 and $M - 1$. In this sense, it is more appropriate to rewrite Equation (4) as,

$$j = \left[-(l+\sigma) + Mq\right] \mod M. \tag{5}$$

Hence, a two-step procedure is suggested to link the topological charge of the scattered field with the OAM and the SAM of the incident field through the fold of the rotational symmetry of the PVL: first using Equation (5) to find the index of the irrep; and then using Equation (2) or Equation (3) to determine the original or the *modular* topological charge. Since the PV at the central region is associated with the *modular* topological charge $l'_s$, its topological charge is bounded by $\left[\frac{M-1}{2}\right]$ as previously discussed. Lastly, it is noted that the rotational symmetry becomes a continuous one, i.e., the $C_\infty$ group, the conservation relation $l_s = l + \sigma$ is exact (see SI S5 for the proof).

To verify the relation in Equation (5), a PVL with eight-fold rotational symmetries is deigned, as shown in Figure 3(a)-(b). The PVL consists of eight spiral slits etched into a 105 nm thick gold film over a silica substrate (the refractive index is 1.45). The inner radius of the eight spiral slits is described as $\rho = \rho_2 + M\varphi\lambda_2/2\pi - (a-1)\lambda_2$, where $\rho_2$ is the initial radius of the spiral, $\lambda_2$ is the width of the gap between two spiral slits, $\varphi$ is the azimuthal angle and $a=1,2…M$. In our design, $\rho_2=3.1\lambda_2$, $\lambda_2=\lambda_{spp}$ ($\lambda_{spp}$ is the wavelength of the generated surface plasmon polaritons on the interface between the PVL and air), and the width of the slit is set as 150nm. The wavelength of the incident light is chosen as 660nm. The dielectric constant of the 105nm thick gold film at 660nm is $-11.82+1.24i$ [34]. The PVL is illuminated by left circularly polarized vortex beams $|\sigma, l\rangle$ with $\sigma$ being 1 and

$l$ ranging from -1 to -8, as shown in Figure 3(a). The incident light comes from the back of the structure. Therefore, there are eight combinations of $(\sigma, l)$: (+1, -1), (+1, -2), (+1, -3), (+1, -4), (+1, -5), (+1, -6), (+1, -7) and (+1, -8). Based on Equations (3) and (5), the eight combinations lead to indices of irreps $j$: 8, 1, 2, 3, 4, 5, 6, 7 and *modular* topological charges $l'_s$: 0, -1, -2, -3, NaN, +3, +2, +1. Here, NaN corresponds to the vortex mode associated with the fourth irrep which is an octopole mode.

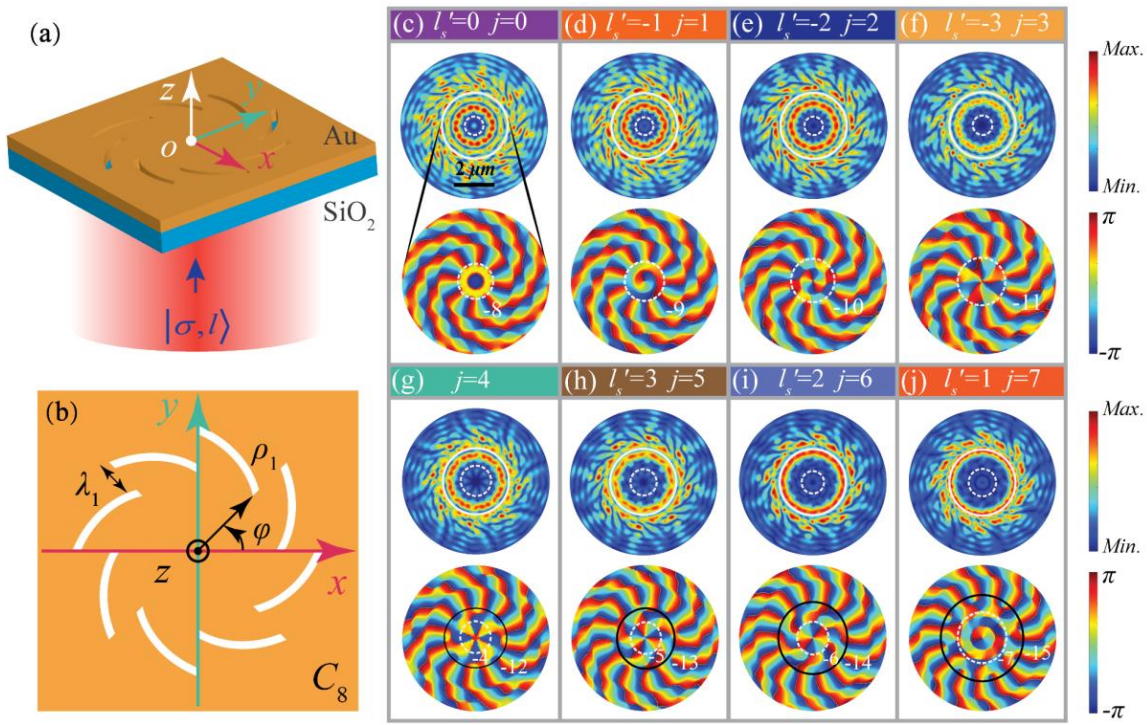

**Figure 3.** Illustration of the PVL with 8-fold rotational symmetries and the simulated magnitude and phase patterns of $E_z$ on a cut above the PVL ($z = 20\ nm$). (a) and (b) illustrate the simulation setup and the top view of the PVL. This incident field is shone from the back the structure. (c)-(j) correspond to the field patterns excited by the incident vortex beam with orbit quantum number $l$ ranging from -1 to -8. In (c) – (j), the zoomed-in phase patterns in the second and the fourth rows are the ones in the central regions

enclosed by the solid white lines in the first and the third rows. The magnitude is coded by the color from blue to red to denote the smallest and the largest values; the phase is coded by the same color to denote the phase variations between $-\pi$ and $+\pi$.

Then, the electric fields excited by the left circularly polarized vortex beams are simulated by using COMSOL Multiphysics. The simulated magnitude and phase patterns of $E_z$ are displayed in Figure 3. In Figure 3(c)-(f) and (h)-(j), it is seen that the fields radiated by the PVL indeed form vortex modes. First, focus on the central regions enclosed by the white dashed lines. In Figure 3(c)-(f) and (h)-(j), the number of full phase variations exhibits 0, -1, -2, -3, +3, +2, +1, respectively. In Figure 3(g), an octopole mode is seen. This observation agrees well with the prediction by first evaluating Equation (5) for the index of irrep $j$ and then using Equation (3) to obtain the *modular* topological charge $l'_s$. Further, in Figure 3(c)-(f) and (h)-(j), it is also seen that, in the regions outside the white dashed lines, more full phase variations exhibit. It is confirmed that the number of these full phase variations is related with the number of full phase variations in the region enclosed by the white dashed lines by $8q$ (see Figure 3), as the discussions around Equation (2). Next, it is seen that no matter how large the incident orbital momentum can be, the *modular* topological charge and thus the topological charge of the central region are always bounded by 3. Lastly, it is worth noting that, when $M = 2$, there is only a trivial vortex mode in the central region according to Equation (2) and (3). As a result, to obtain a non-trivial vortex mode with zero topological charge in the central region, the fold of the rotational symmetries must be at least 3 (see SI S6 for numerical examples).

**Conclusions**

This work provides a symmetry point of view for understanding the spin-orbit coupling in rotationally symmetric PVLs. The symmetries of the PVLs define a correspondence relation between the topological charges that can be radiated by the PVLs and the irreps of the group formed by the symmetries. This correspondence can be many-to-one or one-to-one depending on whether the *ordinary* topological charge or the *modular* topological charge is concerned. It reveals the symmetry origins of the full phase variations in the complex interference patterns in the near field of PVLs. Especially, the *modular* topological charge is bounded by $\left[\frac{M-1}{2}\right]$, where $M$ is the number of rotational symmetries held by the PVL and the square bracket rounds $\frac{M-1}{2}$ to the nearest non-negative integer towards zero. The correspondence relation is complemented by a conservation relation for the spin-orbit coupling in the PVLs. The two relations in combination link the orbit and the spin angular momenta of the incident beam, with the topological charges of the fields scattered by the PVL, through the rotational symmetries of the PVL. It is also worth mentioning that, although the discussions of this work are conducted within the context of the PVLs, the symmetry arguments are generic and hence are not restricted to the optical regime. Therefore, the symmetry perspective provided in the present work serves as an important complement for the physical investigations in the previous works[12,15,16], puts the heuristic arguments in a rigorous mathematical framework, and can be an essential tool for the future engineering of electromagnetic vortices for nanophotonic applications.

**Methods**

**Numerical simulations.** Codes are implemented in MATLAB to generate the irreps, the transformation operators, and the projection operators for cyclic groups. The procedures are as follows: (1) The PVL with the $C_8$ point group symmetries shown in Figure 1(a) is discretized by triangular meshes; (2) The currents flowing in the PVL is approximated by edge-based basis functions[35]; the basis functions are then used to construct matrix representations of the projection operators for the irreps of the $C_8$ point group; and (3) By solving the eigenvalue problem, the "eigencurrents" flowing in the PVL is obtained; and the electric fields radiated by the currents are then calculated and plotted in Figure 1.